\documentclass[reqno]{amsart}

\usepackage{amsmath,amssymb,amsthm,amsfonts}
\usepackage{ulem}
\usepackage{color}
\usepackage{hyperref}

\usepackage{enumerate}
\usepackage{color}

\usepackage{float}
\usepackage{graphicx}

\hypersetup{colorlinks=true,linkcolor=blue,citecolor=blue}
\numberwithin{equation}{section}

\newtheorem{theorem}{Theorem}[section]

\theoremstyle{definition}

\theoremstyle{remark}
\newtheorem{remark}{Remark}

\numberwithin{remark}{section}

\newcommand{\be}{\begin{equation}}
	\newcommand{\ee}{\end{equation}}
\newcommand{\ba}{\begin{equation} \begin{aligned}}
		\newcommand{\ea}{\end{aligned}\end{equation}}
\newcommand{\bes}{\begin{equation*}}
	\newcommand{\ees}{\end{equation*}}

\def\1{{\mathchoice {1\mskip-4mu\mathrm l}      
		{1\mskip-4mu\mathrm l}
		{1\mskip-4.5mu\mathrm l} {1\mskip-5mu\mathrm l}}}

\begin{document}

	\title{New convergence bound for the cluster expansion in canonical ensemble}
	\author{Giuseppe Scola}
	\address{Dipartimento di Matematica e Informatica, Università della Calabria (UNICAL), Cubo 30B,  Via Pietro Bucci, 87036, Arcavacata di Rende (CS), Italy.}
	\email{giuseppe.scola@unical.it}
	\date{}
	
	\maketitle
	
	\begin{abstract} We perform a cluster expansion in the canonical ensemble with periodic boundary conditions, introducing a new choice of polymer activities that differs from the standard one used in \cite{PM,DE1}. This choice leads to an improved bound for the convergence of the cluster expansion, which we compare with the one obtained from \cite{PM,PY}. We also recover the irreducible Mayer coefficients for the thermodynamic free energy. The results presented here can also be applied to the case of zero boundary conditions and to the convergence of correlation expansions as in \cite{DT,DE2}. \\
		
		
	\end{abstract}

	\tableofcontents

	\section{Introduction}
	
	Predicting macroscopic properties from microscopic structures remains a central challenge in statistical mechanics. In the theory of non-ideal gases, an important theoretical and practical contribution was made by J.~E.~and M.~G.~Mayer \cite{MM}, where the authors expressed the pressure $p$ of the system as a power series in the density $\rho$ (virial expansion), viewed as a perturbation of the ideal-gas equation of state at temperature $T$, with convergence proved later (see \cite{R,C}). In \cite{MM}, the thermodynamic pressure is obtained as the infinite-volume limit of the logarithm of the grand canonical partition function, which is a function of the activity of the system. Thus, obtaining an expansion in terms of the thermodynamic density requires passing from activity to density by applying the well-known virial inversion (see, e.g., \cite{J}). In this way, using the standard relation between thermodynamic pressure and thermodynamic free energy (see, e.g.,  \cite[Appendix]{DE1}), one obtains an expansion of the free energy as a power series in the density.
	
	An alternative and more direct way to do this is to compute the thermodynamic free energy as the thermodynamic limit of the canonical partition function, after applying cluster expansion methods directly to the latter. This strategy is adopted, for instance, in \cite{PM,DE1,Mio}, and in the context of density correlation functions in \cite{DT,DE2}.
	
	All the results mentioned above establish the validity of the expansions under suitable assumptions on the activity or on the density of the system. Our aim in this paper is to derive an improved density condition for the convergence of the expansion of the thermodynamic free energy.
	
	In \cite{PM,DE1}, the authors perform a cluster expansion for the canonical partition function of a system of $N$ particles in a box $\Lambda$ interacting via pair potential $V(q_i-q_j)$ at positive inverse temperature $\beta$, that is,
	\[
	Z_\Lambda(N)
	= \frac{1}{N!} \int_{\Lambda^N} dq_1\cdots dq_N\, e^{-\beta \sum_{i,j} V(q_i-q_j)},
	\]
	using a polymer representation such that the activities of polymers with a single element are equal to one. This is achieved by redefining the measure $dq$ as $dq/|\Lambda|$,
	in such a way that the convergence condition for the expansion is of the type 
	\[ C(V,\beta)\frac{N}{|\Lambda|}\le C^*
	\] 
	with $C(V,\beta)$ constant depending on the interactions and the inverse temperature, for some positive constant $C^*$.
	
	In our approach, instead of normalizing with respect to $|\Lambda|$, we normalize the measure $dq$ with $K|\Lambda|$ ($K\ge1$), which again allows polymers with a single element to be present. Thus, $K$ serves as a free parameter that can be optimized to improve the convergence condition for the cluster expansion of $Z_\Lambda(N)$.
	
	Moreover, as emphasized in Remark \ref{Rem_zc}, and following \cite{DT,DE2}, the analysis presented here - carried out for periodic boundary conditions - can also be applied in the case of zero boundary conditions and to prove convergence of multi-body correlation functions as power series in the density.
	
	\subsection{Structure of the paper}
	
	In Section~\ref{sec:model-main} we present the model, state the main result of the paper, and give some useful remarks in which we also compare our convergence condition with the known one.
	
	In Section~\ref{sec:polymer} we prove convergence of the cluster expansion under the new bound. 
	
	Finally, in Section~\ref{sec:irreducible} we analyse the structure of the coefficients in the expansion in order to recover the irreducible Mayer's coefficients \cite{MM}.
	
	\section{Model and main result}\label{sec:model-main}
	
	We consider a system of $N$ interacting particles in the box $\Lambda:=\big(-\frac{L}{2},\frac{L}{2}\big]^d\subset \mathbb{R}^d\;(L>0)$ interacting via a pair potential $V:\mathbb{R}^d\to\mathbb{R}$ at positive inverse temperature $\beta$. Since our interest is in the infinite-volume limit of the free energy, we work with periodic boundary conditions. Denoting a particle configuration by ${\boldsymbol{q}} \equiv \{q_1,\dots,q_N\}$, we assume that $V$ is a \emph{stable} and \emph{tempered} potential, i.e.:
	\begin{itemize}
		\item (\emph{Stability}) there exists $B\ge 0$ such that
		\begin{equation}\label{stability}
			\sum_{1\le i<j\le N}V(q_i-q_j)\ge -BN,	
		\end{equation}		
		for all $N$ and all configurations $q_1,\dots,q_N$;
		\item (\emph{Temperedness}) for all $\beta>0$,
		\begin{equation}\label{C}
			C(\beta)=\int_{\mathbb{R}^d} dx \left(1-e^{-\beta |V(x)|}\right)< \infty.	
		\end{equation}	
	\end{itemize}
	
	Periodic boundary conditions are implemented by tiling $\mathbb{R}^d$ with copies of $\Lambda$ and summing the interaction over all periodic images, that is
	\begin{equation}\label{Vper}
		V^{\rm{per}}(q_i,q_j):=\sum_{n\in\mathbb{Z}^d}	V(q_i-q_j+nL).
	\end{equation}	
	To guarantee the convergence of this series we impose a decay condition on $V$. We say that $V$ is \emph{lower regular} if there exists a decreasing function $\psi:\mathbb{R}^+\to\mathbb{R}^+$ such that $V(x)\ge -\psi(|x|)$ for all $x\in\mathbb{R}^d$, and $\int_0^\infty ds\;s^{d-1}\psi(s)<\infty$. We call $V$ \emph{regular} if it is lower regular and there exists a finite positive real number $r_V$ such that $V(x)\le \psi(|x|)$ whenever $|x|\ge r_V$; this is our additional assumption (see also \cite{FL}).
	
	We define the {\it{canonical partition function}} of the system with periodic boundary conditions by
	\begin{equation}\label{eq:Z-per}
		Z^{\mathrm{per}}_{\Lambda,\beta}(N)
		\;:=\;\frac{1}{N!}\int_{\Lambda^N} dq_1\cdots dq_N\,
		\exp\Bigl(-\beta \sum_{1\le i<j\le N} V^{\mathrm{per}}(q_i,q_j)\Bigr),
	\end{equation}
	and the {\it{thermodynamic free energy}} as the limit of the {\it{finite-volume free energy}}, i.e.
	\begin{equation}\label{eq:f-thermo}
		f_\beta(\rho)\;:=\;\lim_{\substack{N,|\Lambda|\to\infty\\ N/|\Lambda|=\rho}}
		f^{\mathrm{per}}_{\beta,\Lambda}\!\left(\frac{N}{|\Lambda|}\right),
	\end{equation}
	where the finite-volume free energy is given by
	\begin{equation}\label{eq:f-finite}
		f^{\mathrm{per}}_{\beta,\Lambda}\!\left(\frac{N}{|\Lambda|}\right)
		\;:=\;-\frac{1}{\beta|\Lambda|}\log Z^{\mathrm{per}}_{\Lambda,\beta}(N).
	\end{equation}
	
	To prove our theorem we will follow the proofs presented in \cite{PM,PY} so that, for later use, we define
	\begin{equation}\label{eq:F-def}
		F(u) := \max_{a>0} \frac{\log\bigl[1+u(1-e^{-a})\bigr]}{e^{a}\bigl[1+u(1-e^{-a})\bigr]},
	\end{equation}

Our main result is the following.
	
	\begin{theorem}\label{thm:main}
		Given inverse temperature $\beta>0$ and stability constant $B\ge 0$, there exists $K\ge1$, such that for all complex $\rho$ with $|\rho|\le \rho^*$, where 
		\begin{equation}
			\rho^*:=\frac{e^{-\beta B}K}{C(\beta)}F(e^{-\beta B}K),
		\end{equation}
		the limit in \eqref{eq:f-thermo} exists and the thermodynamic free energy is given by
		\begin{equation}\label{eq:TFE-final}
			f_\beta(\rho) \;=\; \frac{1}{\beta}\left\{
			\rho(\log\rho -1) \;-\;\sum_{m\ge 1} \frac{\rho^{m+1}}{m+1}\,\beta_m
			\right\},
		\end{equation}
		where the coefficients $\beta_m$ are the irreducible Mayer coefficients, defined by
		\begin{equation}\label{eq:Mayer-coeff}
			\beta_m \;:=\; \frac{1}{m!}
			\sum_{\substack{g\in \mathcal{B}_{m+1}\\ 1\in V(g)}}
			\int_{(\mathbb{R}^d)^m}
			\prod_{\{i,j\}\in E(g)}\bigl(e^{-\beta V(q_i,q_j)}-1\bigr)\,
			dq_2\cdots dq_{m+1},
			\qquad q_1\equiv 0,
		\end{equation}
		and $\mathcal{B}_{m+1}$ denotes the family of $2$-connected (irreducible) graphs with vertex set $\{1,\dots,m+1\}$.
	\end{theorem}
	
	The proof of Theorem~\ref{thm:main} is divided into two parts. In Section~\ref{sec:polymer} we prove convergence of the cluster expansion for the canonical partition function (that is, of the expansion of the finite-volume free energy $f^{\mathrm{per}}_{\beta,\Lambda}(N/|\Lambda|)$). In Section~\ref{sec:irreducible} we derive formula \eqref{eq:TFE-final} for the thermodynamic free energy.
	
	\begin{remark}\label{rem:comp}[Comparison with known radius of convergence]
		
		As noted in \cite[Remark 2]{PM}, the function $F(x)$ defined in \eqref{eq:F-def} , is an increasing function of $x$ with $F(0)=0,\;F(1)\simeq0.1448$ and $\lim_{x\to\infty}F(x)=e^{-1}$. On the other hand, $a$ is decreasing function of $x$ so that, $a_0\simeq 38.197$, $a_x\simeq 1$ if $x\simeq 5.2691\cdot 10^{-16}$ and $\lim_{x\to \infty} a_x=0$.
		
		Then, comparing the bound $\rho^*$ in Theorem \ref{thm:main} with the one obtained from \cite{PM, PY}, given by:
		\begin{equation}\label{Pbound}
			\rho^*_{1}:=\frac{e^{-\beta B}{F}(e^{-\beta B})}{C(\beta)}	
		\end{equation}	
		with ${F}(u)$ defined in \eqref{eq:F-def}, 
		we have: 		
		\begin{equation}
		\rho^*\ge \rho^*_1,
		\end{equation}	
		if 
		\[KF(e^{-\beta B}K)\ge F(e^{-\beta B}),\]
		which is true for $K\ge 1$.
		
		In particular, if we consider the hard-core case in $\mathbb{R}^d$, i.e. $B=0$ and $C(\beta)=|B_r|$, the volume of the ball of radius $r$ centred at the origin, we get 
		\begin{equation}
			\rho^*=\frac{K}{|B_r|} F(K)\ge\rho^*_1=\frac{1}{|B_r|}F(1),
		\end{equation}
		where $K$ has to be properly chosen (see \eqref{ineq1} in Section \ref{sec:polymer}). In particular, one can prove that, in this case, Theorem \ref{thm:main} is valid if we choose $K\in[1,1.1462]$, so that the maximum value of $\rho^*$ defined in \eqref{rhoStar} is given by $\rho^*= 0.1794 |B_r|^{-1}$ (for $K=1.1462$) while, for $\rho^*_1$ defined in \eqref{Pbound}, we have $\rho^*_1=0.1448|B_r|^{-1}$.
	\end{remark}

	\begin{remark}		\label{Rem_zc}[Zero boundary conditions and correlations] 
		
		The case of zero boundary conditions (no particles outside $\Lambda$) is described by the Hamiltonian $H_\Lambda(\boldsymbol{q}):=\sum_{1\le i<j\le N}V(q_i-q_j)$. Then, as we will see in Section \ref{sec:polymer}, the polymer activities for $|V|=1$ are not affected by the presence of the boundary (see \eqref{eq:zeta-def}), so that the same result as in \cite{DE2} can be recovered. Thus, in this case, the convergence of the cluster expansion with the new activities and the improved radius of convergence can be established. This implies the validity of analogous estimates to those given in \cite[Sections 4,5,6]{DE2}  up to a term of order $|\partial \Lambda|/|\Lambda|$, where $|\partial \Lambda|$ denotes the measure of the boundary of $\Lambda$.
		
		Finally, if we consider correlations as is done in \cite{DT}, proceeding as in Section \ref{sec:irreducible}, it is possible to find similar results with the same cluster structure of  \cite{DT} in the case of p.b.c., with the improved radius of convergence. 	
	\end{remark}

	\section{Polymer model representation and convergence of cluster expansion}\label{sec:polymer}
	
	Following \cite{DE1}, we start by representing the partition function in \eqref{eq:Z-per} as a polymer model. Let $K$ be a positive constant grater than 1.
	We set 
	\begin{equation}\label{eq:Z-free}
		Z^{\mathrm{free}}_{\Lambda,\beta}(N) := \frac{(|\Lambda|K)^N}{N!},
	\end{equation}
	and
	\begin{equation}\label{eq:Z-int}
		Z^{\mathrm{int}}_{\Lambda,\beta}(N)
		:=\int_{\Lambda^N} \lambda(dq_1)\cdots \lambda(dq_N)\,
		\exp\Bigl(-\beta\sum_{1\le i<j\le N} V^{\mathrm{per}}(q_i,q_j)\Bigr),
	\end{equation}
	where
	\begin{equation}\label{eq:lambda-measure}
		\lambda(dq) := dq\,\bigl(|\Lambda|K\bigr)^{-1}.
	\end{equation}
	With these definitions, the partition function \eqref{eq:Z-per} can be written as
	\begin{equation}\label{eq:Z-factor}
		Z_{\Lambda,\beta}(N)
		= Z^{\mathrm{free}}_{\Lambda,\beta}(N)\,Z^{\mathrm{int}}_{\Lambda,\beta}(N).
	\end{equation}
	
	We now express $Z^{\mathrm{int}}_{\Lambda,\beta}(N)$ in terms of a polymer representation. Let
	\begin{equation}\label{eq:fij}
		f_{i,j} \equiv f(q_i,q_j) := e^{-\beta V(q_i,q_j)} - 1.
	\end{equation}
	For $V\subset [N]$, define
	\begin{equation}\label{eq:zeta-tilde}
		\widetilde\zeta(V) :=
		\begin{cases}
			\displaystyle
			\sum_{g\in \mathcal{C}_V}
			\int_{\Lambda^{|V|}}
			\prod_{\{i,j\}\in E(g)} f_{i,j}\,
			\prod_{i\in V} \lambda(dq_i),
			& |V|\ge 2,\\[1.5ex]
			\displaystyle
			\frac{1}{K}, & |V|=1,
		\end{cases}
	\end{equation}
	where $\mathcal{C}_V$ denotes the set of connected graphs with vertex set $V$, and we used the identity $K^{-1} = \int_\Lambda \lambda(dq).$
	
	Given $V,V'\subset [N]$, we say that $V$ and $V'$ are \emph{compatible}, and write $V\sim V'$, if $V\cap V'=\emptyset$. They are \emph{incompatible}, written $V\not\sim V'$, otherwise. Proceeding as in \cite{PM,DE1}, one obtains
	\begin{equation}\label{eq:Zint-poly1}
		\begin{aligned}
			Z^{\mathrm{int}}_{\Lambda,\beta}(N)
			&=\sum_{l\ge 0}\frac{1}{l!}
			\sum_{\substack{V_1,\dots,V_l\subset [N]\\ \{V_i\}_{i=1}^l\ \text{partition of }[N]}}
			\phi(V_1,\dots,V_l)\,\prod_{i=1}^l \widetilde\zeta(V_i)\\
			&=\sum_{l\ge 0}\frac{1}{l!}
			\sum_{\substack{V_1,\dots,V_l\subset [N]\\ |V_i|\ge 1}}
			\phi(V_1,\dots,V_l)\,\prod_{i=1}^l \zeta(V_i),
		\end{aligned}
	\end{equation}
	where
	\begin{equation}\label{eq:zeta-def}
		\zeta(V) :=
		\begin{cases}
			\widetilde\zeta(V), & |V|\ge 2,\\[0.5ex]
			\displaystyle \frac{1}{K}-1, & |V|=1,
		\end{cases}
	\end{equation}
	and
	\begin{equation}\label{eq:phi-def}
		\phi(V_1,\dots,V_l)
		:= \sum_{g\in \mathcal{G}_l}
		\prod_{\{i,j\}\in E(g)} \mathbf{1}_{\{V_i\sim V_j\}},
	\end{equation}
	with $\mathcal{G}_l$ the set of graphs with vertex set $\{1,\dots,l\}$.
	
	In passing from the first to the second line of \eqref{eq:Zint-poly1} we use, for $|V|=1$, the identity $\widetilde\zeta(V) = \zeta(V) + 1$ and expand the resulting product (see \cite[Sec. 3.2]{J1}). 
	
	It is well known that the   logarithm of the hard core polymer gas partition
	function can be written as (see e.g.\ \cite{PS})
	\begin{equation}\label{Polym}
		\log	Z^{\rm{int}}_{\Lambda,\beta}(N)=\sum_{n\ge 1}\frac{1}{n!}\sum_{\substack{V_1,\dots,V_{n}\subset[N]\\ |V_k|\ge 1, \forall k\in[n]}}\phi^T(V_1,\dots,V_{n})\prod_{k=1}^{n}\zeta(V_k),
	\end{equation}	
	with
	\begin{equation}\label{phiT}
		\phi^T(V_1,\dots,V_n):=\sum_{g\in \mathcal{C}_n}\prod_{\{i,j\}\in E(g)}-\mathbf{1}_{\{V\not\sim V'\}}(V_i,V_j),
	\end{equation}	
	and $\mathcal{C}_n$ denotes the family of connected graphs with vertex set $\{1,\dots,n\}$.
	
	We must ensure the convergence of the series in \eqref{Polym}, which is absolutely convergent (see \cite{FP}), when
	\begin{equation}\label{convcond}
		\sup_{i\in[N]}\sum_{\substack{V\subset[N]\;:\;i\in V\\|V|\ge 1}}|\zeta(V)|e^{c|V|}\le e^c-1,
	\end{equation}	
	for some $c>0$.
	
	Since
	\begin{equation}\label{conv1}
		\sup_{i\in[N]}\sum_{\substack{V\subset[N]\;:\;i\in V\\V\ge 1}}|\zeta(V)|\le 1-\frac{1}{K}+\sup_{i\in[N]}\sum_{\substack{V\subset[N]\;:\;i\in V\\|V|\ge 2}}|\zeta(V)|,	
	\end{equation}	
	in order to verify \eqref{convcond}, we need to analyse the second term in the r.h.s. of \eqref{conv1}.

	To do this, following \cite{PM}, we seek a condition on the density $N/|\Lambda|$ such that, for some $a>0$,
	\begin{equation}\label{eq:convergence-cond}
		\sup_{i\in [N]} \sum_{\substack{V\subset [N]:\, i\in V\\ |V|\ge 2}}
		|\zeta(V)|\,e^{a|V|} < e^a - 1.
	\end{equation}
	
	Using the fact that the sum in the l.h.s does not depend on $i\in[N]$ and that $|\zeta(V)|$ depends only on the cardinality of $V$, calling $|V|=n$ and having that
	\begin{equation}
		|\zeta(V)|=|\zeta_n|{N-1\choose n-1},	
	\end{equation}
	condition \eqref{eq:convergence-cond} can be written as
	\begin{equation}\label{eq:convergence-cond1}
		\sum_{n=2}^N {N-1 \choose n-1}|\zeta_n| e^{an}\le e^a-1.
	\end{equation}	
	
	Denoting by $C_\Lambda(\beta)$ the quantity defined in \eqref{C}, where the integral runs over $\Lambda$ instead of $\mathbb{R}^d$, from usual bounds (cf. \cite[Eq.~(2.12)]{PY},  without the factor $1/n!$), one has
	\[
	|\zeta_n|e^{an}\le \frac{e^{(\beta B+a)}}{K}\;n^{n-2}\left(\frac{e^{\beta B+a}C_\Lambda(\beta)}{K|\Lambda|}\right)^{n-1},
	\]
	so that \eqref{eq:convergence-cond1} is satisfied when 
	\[\sum_{n=2}^N \frac{n^{n-2}}{(n-1)!}\left(e^{\beta B+a}C_\Lambda(\beta)\frac{N}{|\Lambda|}\right)^{n-1}\le e^{-\beta B}K\left(1-e^{-a}\right)
	\]
	i.e., calling $\kappa=\left(\frac{e^{\beta B}C_\Lambda(\beta)}{K}\frac{N}{|\Lambda|}\right)^{-1}$, if
	\begin{equation}\label{eq:convergence-cond2}
		\sum_{n\ge 1}\frac{n^{n-1}}{n!}	\left(\frac{e^a}{\kappa}\right)^{n-1}\le 1+e^{-\beta B}K\left(1-e^{-a}\right).
	\end{equation}	
	
	Hence, following \cite{PM}, we define
	\begin{equation}
		G(e^{-\beta B}K):=\min_{a\ge0 }\inf\left\{\kappa\;:\;\sum_{n\ge 1}\frac{n^{n-1}}{n!}\left(\frac{e^a}{\kappa}\right)^{n-1}\le 1+ e^{-\beta B}K(1-e^{-a})\right\},
		\label{K}
	\end{equation}	
	From \cite{PM}, $G(e^{-\beta B}K)$ can be written as 
	\begin{equation}
		G(e^{-\beta B}K)=\min_{a\ge0}\frac{e^a\left[1+ e^{-\beta B}K(1-e^{-a})\right]}{\log[1+ e^{-\beta B}K(1-e^{-a})]}, 		
	\end{equation}	
	so that, we get equation \eqref{eq:convergence-cond2} and hence \eqref{eq:convergence-cond}, hold if
	\[\left(\frac{e^{\beta B}C_\Lambda(\beta)}{K}\frac{N}{|\Lambda|}\right)^{-1}\ge G(e^{-\beta B}K)
	\]
	that is, 
	\begin{equation}\label{rhoStar}
		\frac{N}{|\Lambda|}\le \rho^*:=\frac{K}{e^{\beta B}C_\Lambda(\beta)}F(e^{-\beta B}K)	
	\end{equation}	
	with $F$ defined in \eqref{eq:F-def}.	
	
	Note that if $K=1$, i.e. the polymers with one element are equal to zero, we obviously recover the radius of convergence of \cite{PM}.

	Then, given $\beta,\;B\;K$, denoting by $a^*$ the value of $a$ in \eqref{K} at which the maximum is attained and choosing $c=a^*$, condition \eqref{convcond} is satisfied by \eqref{eq:convergence-cond}, when
	\begin{equation}\label{ineq}
		\left(1-\frac{1}{K}\right)e^{a^*}+\frac{e^{a^*}-1}{e^{a^*}}\le e^{a^*}-1
	\end{equation}	 
	which is true if $K$ is such that 
	\begin{equation}\label{ineq1}
		K\le \left[1-\left(\frac{e^{a^*}-1}{e^{a^*}}\right)^2\right]^{-1}
	\end{equation}	
	where, from \eqref{K}, $a^*\equiv a^*_{\beta,B,K^*}$

	Setting\[g(x):=\left[1-\left(\frac{e^{x}-1}{e^{x}}\right)^2\right]^{-1},\] 
	we have that $g$ is an increasing function of $x$ while $a_x$ is a decreasing function (see Remark \ref{rem:comp}). Then, noting that $e^{-\beta B}K\in(0,K]$, from a numerical analysis of \eqref{ineq1}, we have 
	\begin{equation}
	\max\{K\ge 1\;|\;\eqref{ineq1}\;{\rm{holds}}\;\forall\;\beta>0,B\ge0\}=:K^*=1.1462,
	\label{K*}
	\end{equation}
so that $g(a_{\beta,B,K^*})\simeq1.1463$ (where $a_{\beta,B,K^*}\ge 0.4421$).

\begin{remark}
Given $\beta,B$ such that $e^{-\beta B}$ is \textquotedblleft small enough\textquotedblright, one can choose $K>K^*$, in such a way that the radius of convergence $\rho^*$ defined in \eqref{rhoStar}, evaluated in $K$ is larger than the one evaluated in $K^*$. For example, as seen in Figure \ref{fig:2}, if we choose $K=1.3$ we have that \eqref{ineq1} is true for all $\beta,B$ such that $1.3\cdot e^{-\beta B}\le 0.1099$, i.e. if $\beta B\ge 2.476$. In this case, for the values of $\beta,B$ for which \eqref{ineq1} holds, we get $1.3\cdot F(1.3\cdot e^{-\beta B})\in (0,0.56586]$, while $F(e^{-\beta B})\in(0,0.34507]$ (see Remark \ref{rem:comp}).

Figure \ref{fig:2} corresponds to case of $K=K^*$ defined in \eqref{K*}, and  $\beta>0,\;B\ge 0$.

\begin{figure}[H]
	\centering
	\includegraphics[width=.8\textwidth]{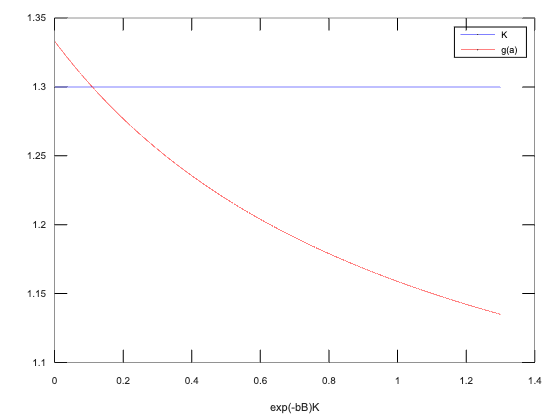}
	\includegraphics[width=.8\textwidth]{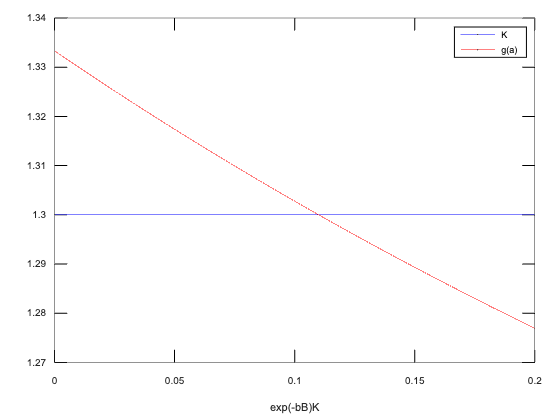}
		\caption{First graph: red line $g(a^*_{\beta,B,K})$, blue line $K=1.3$ with $e^{-\beta B}K\in(0,1.3]$. Second graph: particular with $e^{-\beta B}K=1.3\cdot e^{-\beta B}\in(0,0.2]$.} 
	\label{fig:2}
\end{figure}

\begin{figure}[H]
	\centering
	\includegraphics[width=.8\textwidth]{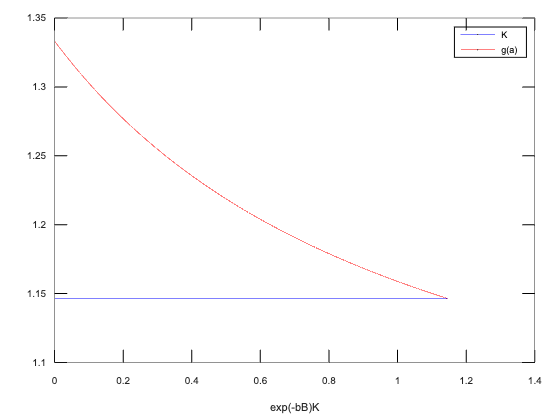}
	\caption{ Red line $g(a^*_{\beta,B,K})$, blue line $K=1.1462$ with $e^{-\beta B}K\in(0,1.1462]$.}
	\label{fig:1}
\end{figure}

\end{remark}

	\section{From cluster coefficients to Mayer's coefficients (proof of \eqref{eq:TFE-final})}\label{sec:irreducible}
	
	To conclude our analysis we organize the structure of the cluster expansion \eqref{Polym} in order to recover the irreducible Mayer's coefficients \eqref{eq:Mayer-coeff}. Starting from \eqref{Polym} and following \cite[Sec. 5]{DE1}, we rewrite the finite-volume free energy defined in \eqref{eq:f-finite} as
	\begin{equation}\label{eq:cluster-structure}
		-\beta\,f^{\mathrm{per}}_{\beta,\Lambda}\!\left(\frac{N}{|\Lambda|}\right)
		= \frac{N}{|\Lambda|}
		\sum_{n\ge 0}\frac{1}{n+1}\,P_{N,|\Lambda|}(n)\,B_{\Lambda,\beta}(n),
	\end{equation}
	where
	\begin{equation}\label{eq:P-def}
		P_{N,|\Lambda|}(n) :=
		\begin{cases}
			\displaystyle
			\frac{(N-1)\cdots (N-n)}{|\Lambda|^n}, & n\ge 1,\\[1ex]
			0, & n=0,
		\end{cases}
	\end{equation}
	and
	\begin{equation}\label{eq:B-Lambda-beta}
		B_{\Lambda,\beta}(n)
		:= \frac{|\Lambda|^n}{n!}
		\sum_{m\ge 1}\frac{1}{m!}
		\sum_{\substack{V_1,\dots,V_m\subset [N]\\ |V_i|\ge 1,\ i\in [m]\\
				\bigl|\cup_{i\in[m]}V_i\bigr| = n+1}}
		\phi^T(V_1,\dots,V_m)\,
		\prod_{i=1}^m \zeta(V_i).
	\end{equation}
	Moreover, $\frac{N}{|\Lambda|}P_{N,|\Lambda|}(n)\longrightarrow \rho^{n+1}$ as
	$|\Lambda|,N\to\infty,\ N/|\Lambda|=\rho.$
	
	\subsection{Free case}\label{subsec:free-case} 
	
	We first consider the contribution with $n=0$ in \eqref{eq:cluster-structure}, i.e.\ the case where no $V$ has more than one element. In this case one finds
	\begin{equation}\label{eq:B0}\begin{split}
			\frac{N}{|\Lambda|}B_{\Lambda,\beta}(0)
			&= \frac{N}{|\Lambda|}
			\sum_{m\ge 1}\frac{(-1)^{m-1}}{m}
			\Bigl(\tfrac{1}{K}-1\Bigr)^{m}\\
			&= \frac{N}{|\Lambda|}\log\Bigl(1+\tfrac{1}{K}-1\Bigr)
			= -\frac{N}{|\Lambda|}\log K.
		\end{split}
	\end{equation}
	On the other hand, from \eqref{eq:Z-free} we have
	\begin{equation}\label{eq:free-partition}
		\frac{1}{|\Lambda|}\log Z^{\mathrm{free}}_{\Lambda,\beta}(N)
		= \frac{N}{|\Lambda|}\log K
		+ \frac{1}{|\Lambda|}\log\frac{|\Lambda|^{N}}{N!},
	\end{equation}
	and, as usual, if $N/|\Lambda|\to \rho$ then
	\begin{equation}\label{eq:ideal-gas}
		\lim_{|\Lambda|,N\to\infty \atop N/|\Lambda|=\rho}
		\frac{1}{|\Lambda|}\log\frac{|\Lambda|^{N}}{N!}
		= \rho(1-\log\rho).
	\end{equation}
	Therefore the first term in \eqref{eq:free-partition} cancels exactly against \eqref{eq:B0}.
	
	\begin{remark}\label{rem:Penrose-tree}
		Recall that \eqref{eq:B0} comes from the fact that, in this case, the only graph in \eqref{phiT} is the complete graph on $n$ vertices. Using the identity (see \cite{PY})
		\begin{equation}\label{eq:Penrose-tree}
			\phi^T(V_1,\dots,V_n)
			= (-1)^{n-1}|\mathcal{T}_R|
			= (-1)^{n-1}\sum_{\tau\in\mathcal{T}_n}\mathbf{1}_{\{\tau\in \mathcal{P}_g(V_1,\dots,V_n)\}},
		\end{equation}
		where $\mathcal{P}_g(V_1,\dots,V_n)$ is the set of Penrose trees of the graph $g\in\mathcal{C}_n$ (with $V_i\not\sim V_j$) rooted at a fixed vertex in $[n]$, and $|\mathcal{T}_R|=(n-1)!$ for the Penrose partition scheme $R$, one recovers \eqref{eq:B0}.
	\end{remark}
	
	\subsection{Interacting case}\label{subsec:interacting}
	
	We now turn to the terms with $n\ge 1$ in \eqref{eq:cluster-structure}. We rewrite \eqref{eq:B-Lambda-beta} as
	\begin{equation}\label{eq:B-Lambda-beta-split}
		\begin{aligned}
			B_{\Lambda,\beta}(n)
			&:= \frac{|\Lambda|^n}{n!}
			\sum_{m\ge 1}\frac{1}{m!}
			\sum_{V_1,\dots,V_m\subseteq[N]}^*
			\phi^T(V_1,\dots,V_m)\,
			\prod_{i=1}^m \zeta(V_i)
			\;+\;R_{\Lambda,\beta}(n),
		\end{aligned}
	\end{equation}
	where the superscript $*$ means that the sum in \eqref{eq:B-Lambda-beta-split} runs over polymers such that 
	\begin{itemize}
		\item $|V_i|\ge 1,$ for all $i\in[m],$
		\item$\bigl|\cup_{i\in[m]}V_i\bigr|=n+1,$
		\item $|V_i\cap V_j|\in\{0,1\},$ for all $i,j\in[m]$,	
	\end{itemize}	
	and, as in \cite{DE1},
	\begin{equation}\label{eq:R-bound}
		|R_{\Lambda,\beta}(n)|\;\le\;\frac{C}{|\Lambda|}.
	\end{equation}
	A direct computation of the algebraic cancellations in
	\begin{equation}\label{eq:hard-part}
		\frac{|\Lambda|^n}{n!}
		\sum_{m\ge 1}\frac{1}{m!}
		\sum_{V_1,\dots,V_m\subset [N]}^*
		\phi^T(V_1,\dots,V_m)\,
		\prod_{i=1}^m \zeta(V_i)
	\end{equation}
	is rather involved, due to the combinatorics generated by the presence of polymers with a single element, weighted by $\bigl(K\bigr)^{-1}-1$.
	
	In order to recover \eqref{eq:TFE-final} we proceed in two steps:
	\begin{enumerate}
		\item \label{I} we explicitly identify the terms converging to the irreducible Mayer's coefficients \eqref{eq:Mayer-coeff};
		\item \label{II} we compare the thermodynamic free energy obtained from \eqref{eq:cluster-structure} with the usual expression, and deduce the cancellation of the remaining coefficients.
	\end{enumerate}
	
	\medskip
	\noindent\textbf{Step \ref{I}.} Consider first the case where there is only one $V$ with $|V|\ge 2$. In this situation one finds (see \cite{PM})
	\begin{equation}\label{eq:oneV-start}
		\begin{aligned}
			&\frac{|\Lambda|^n}{n!}\,\zeta(V)
			\sum_{m\ge 1}\frac{1}{m!}
			\sum_{m_1=1}^m
			\left(\frac{1}{K}-1\right)^{m-1}
			\sum_{V_1,\dots,V_{m-1}\subset [N]}^*
			\phi^T(V_1,\dots,V_m)\\
			&=\frac{|\Lambda|^n}{n!}\,w(V)\,K^{-(n+1)}
			\sum_{m\ge 1}\frac{1}{(m-1)!}
			\Bigl(1-\tfrac{1}{K}\Bigr)^{m-1}
			\sum_{\tau\in\mathcal{T}_m}
			\sum_{V_1,\dots,V_{m-1}\subset [N]}^{**}
			\mathbf{1}_{\{\tau\in \mathcal{P}_g(V_1,\dots,V_n)\}},
		\end{aligned}
	\end{equation}
	where the superscript $**$ means that
	\begin{itemize}
		\item  $|V_i|=1,$ for all $i\in[m-1],$
		\item $\bigl|\cup_{i=1}^{m-1}V_i\cup V\bigr|=n+1,$
		\item $|V_i\cap V_j|\in\{0,1\},$ for all $i,j\in[m-1],$
		\item $|V_i\cap V|=1,$ for all $i\in[m-1],$
	\end{itemize}
	in the second equality we used \eqref{eq:Penrose-tree} and
	\begin{equation}\label{eq:wV}
		w(V) := \sum_{g\in \mathcal{C}_V}
		\int_{\Lambda^{V}}
		\prod_{\{i,j\}\in E(g)} f_{i,j}\,
		\prod_{i\in V}\frac{dq_i}{|\Lambda|}.
	\end{equation}
	Denoting with $|V|=n+1$, we get
	\begin{equation}\label{eq:general-V}
		\begin{aligned}
			&\frac{|\Lambda|^n}{n!}\,w(V)\,K^{-(n+1)}
			\sum_{m\ge 1}\frac{1}{(m-1)!}
			\Bigl(1-\tfrac{1}{K}\Bigr)^{m-1}\times\\
			&\hspace{3.2cm}\times
			\sum_{\substack{m_1,\dots,m_{n+1}\ge 0\\ \sum_{i=1}^{n+1} m_i = m-1}}
			\binom{m-1}{m_1,\dots,m_{n+1}}
			\prod_{i=1}^{n+1} m_i!.
		\end{aligned}
	\end{equation}
	Using now that 
	\begin{equation}
		\begin{aligned}
			&\sum_{m\ge 1}\frac{1}{(m-1)!}
			\Bigl(1-\tfrac{1}{K}\Bigr)^{m-1}\sum_{\substack{m_1,\dots,m_{n+1}\ge 0\\ \sum_{i=1}^{n+1} m_i = m-1}}
			\binom{m-1}{m_1,\dots,m_{n+1}}
			\prod_{i=1}^{n+1} m_i!\\
			&=\frac{1}{n!}\left[\frac{d^n}{dx^n}x^n\sum_{m\ge1}x^m\right]_{x=1-\frac{1}{K_\Lambda(\beta,B)}}=\frac{1}{n!}\left[\frac{n!}{(1-x)^{n+1}}\right]_{x=1-\frac{1}{K_\Lambda(\beta,B)}}\\
			&=K^{n+1},	
		\end{aligned}
	\end{equation}	
	\eqref{eq:general-V} reduces to
	\[
	\frac{|\Lambda|^n}{n!}\,w(V).
	\]
	Arguing as in \cite{DE1}, if we now introduce the polymer activity
	\begin{equation}\label{eq:w-star}
		w^\ast(V)
		:= \sum_{g\in \mathcal{B}_V}
		\int_{\Lambda^{|V|}}
		\prod_{\{i,j\}\in E(g)} f_{i,j}\,
		\prod_{i\in V}\frac{dq_i}{|\Lambda|},
	\end{equation}
	we find that
	\begin{equation}\label{eq:beta-m-limit}
		\frac{|\Lambda|^m}{m!}\,w^\ast(V)\Big|_{|V|=m+1}
		\longrightarrow \beta_m,
		\qquad |\Lambda|\to\infty,
	\end{equation}
	with $\beta_m$ given in \eqref{eq:Mayer-coeff}.	
	
	\medskip
	\noindent\textbf{Step \ref{II}.}  For the general case of $k$ (at least) two by two incompatible polymers with more than one element we proceed as follows. Let us fix $k$ incompatible polymers with more than one particle and denote them by $V_1,\dots,V_k$  so that $|\bigcup_{i=1}^k V_i|=n+1$. 
	Moreover, given $1\le m_i\le m, i\in[n+1]$, we call  $V_{m_i}$ the set of overlapping polymers with one element, i.e.,

	$$V_{m_i}:=
	\bigg\{\{V_j\}_{j\in J}\;: |J|=m_i,\;|V_j|=1,\;V_j\cap V_{j'}\ne \emptyset, \forall\;j, j'\in J\bigg\},\;m_i\in[m]$$

	Then the corresponding term in \eqref{eq:B-Lambda-beta-split}, is given by
	\begin{equation}\label{genV}
		\begin{split}
			&(-1)^{k-1}\frac{|\Lambda|^{n}}{n!}\frac{\prod_{i=1}^{k}w(V_i)}{K^{n+1+l}}\sum_{m\ge k}\frac{1}{m!}{m\choose k}\left(1-\frac{1}{K}\right)^{m-k}\times\\
			&\hspace{3.5cm}\times	\sum_{\substack{m_1,\dots,m_{n+1}\ge 0\\ \sum m_i=m-k}}{m-k\choose m_1\dots m_{n+1}}\varphi\bigg((V_{m_i})_{i\in[n+1]};(V_{j})_{j\in[k]}\bigg)
		\end{split}
	\end{equation}	
	with
	\begin{equation}\label{BruttaPhi}\begin{split}
			&\varphi\bigg((V_{m_i})_{i\in[n+1]};(V_{j})_{j\in[k]}\bigg)\\
			&\quad:=\sum_{j=1}^{k}\prod_{\ell=1}^{n
				+1}\mathbf{1}_{L^j\left(\{V_s\}_{s\in[m-k]},\{V_l\}_{l\in[k]}\right)}(V_{m_\ell})(m_{\ell}+(j-1))!
		\end{split}	
	\end{equation}	 
	where, denoting with $K((V_I),V_J)$ the complete graph with vertex in $I\cup J$ for some $I\subset[m-k],J\subset[m]\setminus[m-(k+1)]$, we defined
	\begin{equation}\begin{split}
			&L^j\left(\{V_s\}_{s\in[m-k]},\{V_l\}_{l\in[k]}\right)\\
			&:=\{V_s\;:\;s\in[m-k],\; V_{s}\;{\rm{vertex\;of}}\;K((V_I),(V_J)),\; J\subset [k],\;|J|=j\}.
		\end{split}
	\end{equation}	
	
	In \eqref{BruttaPhi}, we count the number of Penrose trees that we can construct from polymers labelled in $[m-k]$ (with one element), overlapping with $j$ polymers with more than one element, as $j$ ranges from 1 to $k$.
	
	Hence, using \eqref{eq:B0}, \eqref{eq:free-partition}, \eqref{eq:B-Lambda-beta-split}, \eqref{eq:beta-m-limit} and \eqref{genV}, the thermodynamic limit of \eqref{eq:cluster-structure} is given by 
	\begin{equation}\label{TFEMio}
		f_\beta(\rho)=\frac{1}{\beta}\Bigg\{\rho(\log\rho-1)-\sum_{m\ge 1}\frac{\rho^{m+1}}{m+1}\beta_m-\sum_{m\ge 2}\frac{\rho^{m+1}}{m+1}B_\beta(m)	\Bigg\}
	\end{equation}	
	where we defined
	\begin{equation}\label{TDCMio}\begin{split}
			B_\beta(m)&:=\frac{1}{m!}\sum_{n\ge2}\frac{1}{n!}\sum_{k=1}^n{n\choose k}(-1)^{k-1}\frac{\prod_{i=1}^kw(V_i)}{K^{m+k+1}}\sum_{\ell\ge k}\left(1-\frac{1}{K}\right)^{\ell-k}\times\\
			&\;\times\sum_{\substack{\ell_1,\dots,\ell_{m+1}\ge 0\\\sum \ell_i=\ell-k}}{\ell-k \choose \ell_1\dots \ell_{m+1}}\phi\bigg((V_{\ell_i})_{i\in[m+1]};(V_{j})_{j\in[k]}\bigg).
		\end{split}
	\end{equation}	
	with polymers $V$'s so that $V\subseteq[m].$

	Comparing \eqref{TFEMio}, with the thermodynamic free energy obtained by \cite{PM}, given by
	\begin{equation}\label{TFED}
		f_\beta(\rho)=\frac{1}{\beta}\left\{\rho(\log\rho-1)-\sum_{m\ge 1}\frac{\rho^{m+1}}{m+1}\beta_m\right\}.
	\end{equation}	
	using the notation given in Remark \ref{rem:comp},
	from \eqref{TFEMio} and \eqref{TFED}, we get
	\begin{equation}
		\sum_{m\ge 4}\frac{\rho^{m+1}}{m+1}B_\beta(m)=0	
	\end{equation}	
	for all $\rho\in(0,\rho^*]\cap(0,\rho^*_1]$, with $\rho^*,\rho^*_1$ given by \eqref{rhoStar},  \eqref{Pbound}, which implies 
	\begin{equation}
		B_\beta(m)=0	
	\end{equation}	
	for all $m\ge 2$ and hence the validity of \eqref{eq:TFE-final}, also when $\rho^*>\rho^*_1$, since $B_\beta(m)$ is independent of $\rho$.

	\bibliographystyle{plain}
	\bibliography{biblioNuovo1}

\end{document}